\documentclass[12pt]{article}
\usepackage{amssymb}
\usepackage{latexsym}
\usepackage{amsmath}
\usepackage{graphicx}
\usepackage{listings}
\usepackage{cite}
\usepackage{authblk}
\usepackage[utf8]{inputenc}
\usepackage{esdiff}
\usepackage{mathtools}
\usepackage{xcolor}
\def\De{\Delta}
\def\th{\theta}
\def\vp{\varphi}

\def\la{\lambda}
\def\ka{\kappa}
\def\om{\omega}
\def\al{\alpha}
\def\be{\beta}

\def\ga{\gamma}
\def\vk{\varkappa}
\def\ve{\varepsilon}
\newcommand{\cmmnt}[1]{}

\title{\Large{Higher Spin Mode Stability for  STU Black Hole Backgrounds}}
\author[1,2,3]{\small {M. Cveti\v{c}\thanks{e-mail: cvetic@physics.upenn.edu}}}  
\author[1,4]{\small {M. A. Liao\thanks{e-mail: matheusalvesliao@gmail.com}}}
\author[1,5]{\small {M. M. Stetsko\thanks{e-mail: mstetsko@gmail.com, mstetsko@upenn.edu}}}
\affil[1]{Department of Physics and Astronomy, University of Pennsylvania, Philadelphia, PA, 19104, USA}
\affil[2]{Department of Mathematics, University of Pennsylvania, Philadelphia, PA, 19104, USA}
\affil[3]{Center for Applied Mathematics and Theoretical Physics, University of Maribor, Maribor, Slovenia}
\affil[4]{Departamento de Física, Universidade Federal da Paraíba, 58051-970 João Pessoa, PB, Brazil}
\affil[5]{Department for Theoretical Physics, Ivan Franko National University of Lviv, Lviv, UA-79005, Ukraine}
\begin{document}
\vspace*{-2cm}
\begin{flushright}
    \texttt{UPR-1327-T}
\end{flushright}
{\let\newpage\relax\maketitle}
\maketitle
\begin{abstract}
	This work studies mode stability of an STU black hole with four pairwise equal $\rm{U(1)}$ charges in four spacetime dimensions. We investigate bosonic perturbations for probe fields of different spins through the transformation technique devised by Whiting in 1989. Finally, we introduce connection relations inspired by the work of ~Duzta\c{s} (2016) to prove the absence of unstable modes that solve the torsion-modified Dirac equation appropriate for this black hole background.
\end{abstract}

\section{Introduction}
The stability of rotating black holes has been the subject of active research in the field of gravitation for many years because of its fundamental importance. In a series of recent papers (see~\cite{Kerr_stability} and references therein) the stability of Kerr black holes has been conclusively proved, at least for sufficiently small angular momentum. This demonstration marked an important triumph of General Relativity, as it proves consistency of the theory with the observed prevalence in nature of black holes of this kind. An important first step in investigations of black hole stability is to prove the absence of exponentially increasing modes in the master equation, derived by Teukolsky for the Kerr metric~\cite{Teukolsky_ApJ1973}. Mode stability for the aforementioned solution was proved by Whiting over three decades ago~\cite{Whiting_JMP89} by virtue of of transformations that map the physical modes into functions which may be of the perturbation equations from an auxiliary physical space. The metric implied by the wave equation in that space does not give rise to an ergoregion, thus evading the difficulties that arise from the presence of that region in spacetime. We also point out that the so-called superradiance instability caused by the ergoregion of a rotating black hole takes place for fields (particles) of an integer spin, but it does not occur in the fermionic case.

The present work aims to show mode stability in the pair-wise equal charged STU black hole background. Such black holes have been previously investigated in Refs~\cite{Cvetic_PRL20,Cvetic_PRD22, NPB717, Cvetic_JHEP,Cvetic_JCAP17, Cvetic_arxiv2023, Cvetic_JHEP12 }  among several others. References~\cite{Cvetic_PRL20,Cvetic_PRD22}, in particular, have successfully made use of Whiting's procedure to demonstrate mode stability for massless scalars ($s=0$) in STU black holes of four and five dimensions respectively. Separability of the torsion modified Dirac equation that we shall use to investigate spin $\pm 1/2$ perturbations of the metric has been proved in a recent work by two of the authors~\cite{Cvetic_arxiv2023}. As far as we know proper perturbations of the considered black hole space-time or field equations for spin $s>1/2$ at this background have not been studied yet, but taking into account the wave equations for $s=0$ and $s=1/2$, we have conjectured equations which can be treated as the generalization of the decoupled Teukolsky equations for the case of the pair-wise equal charged geometry \cite{Cvetic_arxiv2023}. Here we show that instead of the pair of decoupled equations a unique equation can be written and the former follows from the latter after separation of variables. Having this equation we consider perturbations of both bosonic and fermionic nature. The former ones will be studied through the implementation of Whiting's procedure, as was done in the aforementioned papers. Fermionic perturbations, however, require a separate treatment. This necessity stems from the form of Whiting's differential transformations which, as we shall show, are not appropriate for half-integer spin. To circumvent this difficulty, we will use connection relations similar to those introduced in~\cite{Duztas_PRD16} in their study of mode stability for neutrino perturbations.  As shall be explained in details below, this demonstration is based on the assumption that  the expansion coefficient $Y^{(in)}$, relating to ingoing modes at infinity, can be treated as an analytic function of $a$ and $\omega$, with a well-defined nonrotating limit as $a\to 0$. As we will show, these connection relations, which follow from the field equations and separability of the theory, are not consistent with the presence of unstable modes, thus allowing us to rule out their existence.

The paper is structured as follows: after this introductory section, we present, in Section 2, Teukolsky-like equations which are meant to describe perturbations in the metric for different values of $s$, obtained by a rather straightforward extrapolation of the work of Teukolsky~\cite{Teukolsky_ApJ1973} to the STU metric, which we check for the cases $s=0$ and $s=1/2$. In Section 3, we use these equations to perform an analysis with the use of Whiting's transformations, which allow for the introduction of an energy-like functional with a positive integrand, which we may thus use to prove that perturbations are bounded. Although this procedure seems to work without explicit specification of $s$, which appears as parameter in the perturbation equations, we notice that one of Whiting's transformations involves a power of a differential operator, and this power is not integer for fermions. Thus, on Section 4 we try a different route by following a procedure first introduced in~\cite{Duztas_PRD16} to derive connection relations which contradicts the possibility of unstable modes. Finally, we conclude in Section 5, summarizing what has been done throughout the paper and pointing out perspectives meant for further investigations.
\section{Teukolsky equation for a rotating STU black hole}

In this work, we investigate small perturbations of an STU black hole~\cite{Cvetic_PRD96}. We consider only the pair-wise equal charges case, for which separability of the torsion-modified Dirac equation has been demonstrated recently~\cite{Cvetic_arxiv2023}. The metric in the pair-wise equal scenario is of the form \cite{NPB717,Cvetic_PRD96}:
 \begin{equation}
\begin{split}\
 ds^2= -\frac{X}{\De^{1/2}_{0}}\left(dt-a\sin^2{\th}d\vp\right)^2+\De^{1/2}_{0}\left(\frac{dr^2}{X}+d\th^2\right)+&\\\frac{\sin^2{\th}}{\De^{1/2}_0}\left(adt-\left((r+2ms^2_1)(r+2ms^2_2)+a^2\right)d\vp\right)^2 &,\label{BL_pair}
\end{split}
 \end{equation}
 where $X=r^2-2mr+a^2$, $\Delta^{1/2}_{0}=(r+2ms^2_1)(r+2ms^2_{2})+a^2$, while  $s_i=\sinh{\delta_i}, i=1,2$ are defined in terms of the $\rm{U(1)}$ charges $\delta_i$, following the notational conventions of Ref.~\cite{Cvetic_arxiv2023}, we also refer reader to \cite{NPB717, Cvetic_PRD96} for the explicit form of the Lagrangian for the considered model and explicit solutions for all the fields.  
 
 Gravitational perturbations of Kerr metric are uniquely described by the well-known Teukolsky master equation. In addition to these perturbations, which are purely bosonic by nature and correspond to spin $s=\pm 1,\pm 2$, the master equation also allows one to study the massless  scalar ($s=0$) and fermionic ($s=\pm1/2$) cases, as pointed out by Teukolsky himself \cite{Teukolsky_ApJ1973}. Moreover, even spin $s=\pm3/2$ can be examined. Teukolsky equation was crucial in the proof of separability for perturbations of arbitrary spin. As far as we know, gravitational perturbations on a rotating STU black hole background have not been studied yet. In any case, it is tempting to have a corresponding analog of the Teukolsky equation, since it provides a general framework in which gravitational perturbations and massless fields, as well as separability, may be investigated. 

In a recent work by two of the authors \cite{Cvetic_arxiv2023}, it was demonstrated that a torsion-modified Dirac equation on a rotating pair-wise equal charges STU black hole background \cite{Cvetic_PRD96,NPB717} is separable. We note that separability of a massless scalar equation was proven a decade ago for more general setting of four nonequal charges \cite{Cvetic_JHEP12}. The decoupled (separated) equations of motion for spin $s=0,\pm 1/2$  can be written in the form\cite{Cvetic_arxiv2023}:  
\begin{gather}
\label{mast_1}\frac{1}{\sin{\theta}}\frac{\partial}{\partial\theta}\left(\sin{\theta}\frac{\partial {\cal S}_{s}}{\partial \theta}\right)+\left(a^2\omega^2\cos^2{\theta}+2sa\omega\cos{\theta}-\frac{k^2+s^2+2ks\cos{\theta}}{\sin^2{\theta}}+E_{k,s}\right){\cal S}_{s}=0,\\\label{mast_2}
X^{-s}\frac{\partial}{\partial r}\left(X^{s+1}\frac{\partial {\cal R}_{s}}{\partial r}\right)+\left(\frac{K^2(r)-2is(r-m)K(r)}{X}+2is\omega F'(r)-\lambda_{k,s}\right){\cal R}_{s}=0,
\end{gather}
where $K(r)=\om F(r)+ka$, $F(r)=r_1r_2+a^2$, $r_i=r+2ms^2_i$, $i=1,2$, $\lambda_{k,s}=E_{k,s}+a^2\omega^2+2a\omega k-s(s+1)$ and here $\om$ and
 $k$ are the angular frequency and magnetic quantum number respectively. The functions ${\cal S}_{s}(\th)$ and ${\cal R}_{s}(r)$ in the above equations are the angular and radial wave functions respectively, while $\la_{k,s}$ is the separation constant.
 
It is tempting to construct a generalization of Teukolsky master equation for the considered background (\ref{BL_pair}). Fortunately enough, it can be conjectured easily, and the resulting equation takes the form:
\begin{multline}\label{teukolsky_eq1}
\left(\frac{F^2}{X}-a^2\sin^2{\th}\right)\frac{\partial^2\Psi_s}{\partial t^2}-\frac{2a(X-F)}{X}\frac{\partial^2\Psi_s}{\partial t\partial \vp}+\left(\frac{a^2}{X}-\frac{1}{\sin^2{\th}}\right)\frac{\partial^2\Psi_s}{\partial \vp^2}-\\X^{-s}\frac{\partial}{\partial r}\left(X^{s+1}\frac{\partial \Psi_s}{\partial r}\right)-\frac{1}{\sin{\th}}\frac{\partial}{\partial \th}\left(\sin{\th}\frac{\partial \Psi_s}{\partial \th}\right)+2s\left(\frac{a(r-m)}{X}-i\frac{\cos{\th}}{\sin^{2}{\th}}\right)\frac{\partial \Psi_s}{\partial\vp}\\+2s\left(\frac{(r-m)F}{X}-r_1-r_2-ia\cos{\th}\right)\frac{\partial \Psi_s}{\partial t}+\left(s^2\cot^2{\th}-s\right)\Psi_s=0,
\end{multline}
where $\Psi_s$ is the Teukolsky wave function for spin $s$, and $r_i=r+2ms^2_i$ are ``shifted'' radial coordinates. If $\Psi_s$ is assumed to be in the form:
\begin{equation}\label{teukolsky_wf}
\Psi_s(t,r,\th,\vp)=e^{i(\om t+k\vp)}{\cal R}_s(r){\cal S}_{s}(\th),
\end{equation}
the conjectured Teukolsky equation (\ref{teukolsky_eq1}) can be easily decoupled to the system of equations (\ref{mast_1})-(\ref{mast_2}) with $\la_{k,s}$ as the separation constant. It is worth noting that contribution of nonzero spin into the Teukolsky equation (\ref{teukolsky_eq1}) is given by the terms proportional to $s$.

Equation (\ref{teukolsky_eq1}) may be rewritten in a more compact form, which might be more convenient for practical applications. To obtain it, we assume that the wave function $\Psi_s$ (or rather its radial part $R_s$) can be chosen in the form:
\begin{equation}\label{wf_mod}
\Psi_s=X^{-s/2}\bar{\Psi}_s, \quad \Leftrightarrow \quad {\cal R}_s(r)=X^{-s/2}\bar{{\cal R}}_s(r).
\end{equation} 
Finally, the above relation allows one to rewrite the Teukolsky equation (\ref{teukolsky_eq1}) in the following way:
\begin{multline}\label{teuk_eq2}
-\frac{\partial}{\partial r}\left(X\frac{\partial \bar{\Psi}_{s}}{\partial r}\right)+\frac{1}{X}\left(F\frac{\partial}{\partial t}+a\frac{\partial}{\partial \vp}+s(r-m)\right)^2\bar{\Psi}_{s}-\frac{1}{\sin{\th}}\frac{\partial}{\partial\th}\left(\sin{\th}\frac{\partial\bar{\Psi}_s}{\partial\th}\right)-\\\frac{1}{\sin^2{\th}}\left(a\sin^2{\th}\frac{\partial}{\partial t}+\frac{\partial}{\partial \vp}+isa\cos{\th}\right)^2\bar{\Psi}_{s}-2s\left(r_1+r_2+2ia\cos{\th}\right)\frac{\partial\bar{\Psi}_s}{\partial t}=0.
\end{multline} 
We note that both forms of the conjectured Teukolsky equation (\ref{teukolsky_eq1}) and (\ref{teuk_eq2}) strictly speaking are applicable to spins $s=0,\pm1/2$ only. To check its applicability to higher spin fields, corresponding field equations should be derived and studied. In our case we assume that there are specific probe fields minimally coupled to gravity, but we expect that at least the most important features of the equations are properly accounted by the equations (\ref{teukolsky_eq1}) and (\ref{teuk_eq2}).

\section{Stability of Teukolsky equation solutions}

Teukolsky equation showed its advantage at the attempt to solve a quite general puzzling question, namely whether there were unstable solutions (modes). For the Kerr black hole the matter is more subtle due to the presence of an ergoregion outside the black hole. Whiting proposed a procedure which allowed to overcome this difficulty, allowing him to demonstrate the absence of unstable modes of the Teukolsky equation for the Kerr black hole \cite{Whiting_JMP89}. In this work we will show the absence of unstable modes (of both bosonic and fermionic nature) for a rotating STU black hole with pair-wise equal charges. For convenience, we work with Eq.~(\ref{teuk_eq2}). 

Following the notation of \cite{Cvetic_PRL20} and \cite{Cvetic_PRD22}, we introduce new variables instead of the radial $r$ and angular $\th$ coordinates:
\begin{equation}\label{transf_1}
x=\frac{r-r_{-}}{r_{+}-r_{-}}, \quad y=\frac{1}{2}\left(1-\cos{\th}\right).
\end{equation}
We point out here that the same transformation of coordinates was used in our earlier paper \cite{Cvetic_arxiv2023} to rewrite the corresponding equations for the  wave function components in a standard form of Heun equation. However, the aforementioned paper only examined the $s=\pm 1/2$ cases. Now, we note that the radial equation (\ref{mast_2}) can be rewritten as follows:
\begin{equation}\label{rad_e1}
{\cal X}''_{s}+\left(-\tilde{\al}^2+\frac{\tilde{\al}\tilde{\ka}+\tilde{\la}+\frac{1}{2}\tilde{\ka}^2}{x}+\frac{\frac{1}{4}-\tilde{\be}^2}{x^2}+\frac{\tilde{\al}\tilde{\ka}-\tilde{\la}-\frac{1}{2}\tilde{\ka}^2}{x-1}+\frac{\frac{1}{4}-\tilde{\ga}^2}{(x-1)^2}\right){\cal X}_{s}=0,
\end{equation}
where 
\begin{gather}
\nonumber\tilde{\al}=i\om\vk,\quad \tilde{\be}=\frac{s}{2}-\frac{i}{\vk}\left(\eta r_{-}+\xi\right),\quad \tilde{\ga}=\frac{s}{2}+\frac{i}{\vk}\left(\eta r_{+}+\xi\right), \quad \tilde{\ka}=s-i\eta, \\ \label{coeff_1}\tilde{\la}=\frac{1}{2}+\tilde{\al}\left(\tilde{\ga}-\tilde{\be}\right)-\frac{1}{2}\left(\tilde{\ga}-\tilde{\be}\right)^2+\la_{k,s}+\nu, \quad \nu=s,
\end{gather}
and here $\eta=2m\om(1+s^2_1+s^2_2)$, $\xi=4\om m^2s^2_{1}s^2_{2}+ka$, $\vk=r_{+}-r_{-}$. Here we have performed the transformation ${\cal X}_s=(x(x-1))^{1/2}\bar{{\cal R}}_s$ on the radial wave function. We point out that for the particular case $s_1=s_2=0$ the relations (\ref{coeff_1}) are reduced to corresponding parameters derived for standard Kerr solution \cite{Whiting_JMP89}. For the scalar field ($s=0$) the derived relations (\ref{coeff_1}) are in agreement with coefficients, obtained in \cite{Cvetic_PRL20} for the particular case of pairwise equal charges.

For the angular part we perform the transformation ${\cal Y}_s=(y(1-y))^{1/2}{\cal S}_s$ and write:
\begin{equation}\label{ang_e1}
{\cal Y}''_{s}+\left(-{\al}^2+\frac{\al\ka+\la+\frac{1}{2}\ka^2}{y}+\frac{\frac{1}{4}-\be^2}{y^2}+\frac{\al\ka-\la-\frac{1}{2}\ka^2}{y-1}+\frac{\frac{1}{4}-\ga^2}{(y-1)^2}\right){\cal Y}_{s}=0,
\end{equation}
and the coefficients in this equation are given by
\begin{gather}
\nonumber \al=2a\om, \quad \be=\frac{s+k}{2}, \quad \ga=\frac{s-k}{2},\\ \ka=s, \quad \la=\frac{1}{2}+\al(\ga-\be)-\frac{1}{2}(\ga-\be)^2+\la_{k,s}+s.
\end{gather}
Equations (\ref{rad_e1}) and (\ref{ang_e1}) are much more convenient for the computations that will follow. 

Following the Whiting's procedure \cite{Whiting_JMP89} we introduce a new function $\bar{h}(x)$ defined by the integral transformation:
\begin{equation}\label{int_transf}
\bar{h}_s(x)=e^{\hat{\al}x}x^{\hat{\be}}(x-1)^{\hat{\ga}}\int^{+\infty}_{1}e^{2\tilde{\al}xz}e^{-\tilde{\al}z}z^{-\frac{1}{2}-\tilde{\be}}(z-1)^{-\frac{1}{2}-\tilde{\ga}}{\cal X}_s(z)dz,
\end{equation}
where
\begin{gather}
\hat{\al}=-2\tilde{\al}, \quad \hat{\be}=-\frac{1}{2}\left(\tilde{\be}+\tilde{\ga}+\tilde{\ka}\right), \quad \hat{\ga}=-\frac{1}{2}\left(\tilde{\be}+\tilde{\ga}-\tilde{\ka}\right).
\end{gather}
The transformed wave function $\bar{h}_s(x)$ solves the following equation:
\begin{equation}\label{rad_it}
x(x-1)\bar{h}''_{s}+(2x-1)\bar{h}'_s+\left(4\om(\eta m+\xi)x-\la_{k,s}-s+\om^2\vk^2x(x-1)-s^2\frac{x-1}{x}-\eta^2\frac{x}{x-1}\right)\bar{h}_s=0.
\end{equation}
To transform the angular equation (\ref{ang_e1}) a differential transformation should be applied \cite{Whiting_JMP89}. Let us briefly describe the key points of this transformation. It was shown that there is a relation between ${\cal Y}_s(y)$ and a new function $v_s(y)$  which satisfies an equation with the same structure as (\ref{ang_e1}). Namely, if for chosen values of parameters $\ve, \ve', \ve''$ ($\ve, \ve', \ve''=\pm 1$) the parameter $n$, defined as
\begin{equation}
n=\ve''\ga+\ve'\be+\ve\ka
\end{equation}
takes integer values, then the function 
\begin{equation}
v_s(y)=e^{\bar{\al}y}y^{\bar{\be}}(1-y)^{\bar{\ga}}\left(\frac{\partial}{\partial y}\right)^n e^{\ve\al y}y^{\ve'\be}(1-y)^{\ve''\ga}{\cal Y}_s(y)
\end{equation}
is a solution of an equation similar to (\ref{ang_e1}), if barred  parameters take the form:
\begin{equation}
\bar{\al}=-\ve\al,\quad \bar{\be}=\frac{1}{2}(1+n)-\ve'\be,\quad \bar{\ga}=\frac{1}{2}(1+n)-\ve''\ga.
\end{equation}
In order to achieve our goal, only two options for the order of derivative $n$ are possible, namely $n=|s-m|$ and $n=|s+m|$. Taking the latter one, we arrive at the equation:
\begin{equation}
v''_{s}(y)+\left(-4a^2\om^2+\frac{\la_{k,s}+s+\frac{1}{2}-4a\om k}{x}-\frac{\la_{k,s}+s+\frac{1}{2}}{x-1}+\frac{1/4}{x^2}+\frac{1/4-s^2}{(x-1)^2}\right)v_s(y)=0.
\end{equation}
Introducing new function $\bar{v}_s(y)=\sqrt{y(1-y)}v_s(y)$, we rewrite it in the form:
\begin{equation}\label{ang_dt}
\left(x(1-x)\partial^2_{x}+(1-2x)\partial_x-4a^2\om^2x(1-x)+4a\om k(x-1)+\la_{k,s}+s(s+1)-\frac{s^2}{1-x}\right)\bar{v}_s(y)=0.
\end{equation}

The key step towards proving the stability of Teukolsky equation solutions or the absence of unstable modes is the so-called ``unseparation of variables''. Since equations (\ref{rad_it}) and (\ref{ang_dt}) have the same separation constant $\la_{k,s}$, which comes from the original Teukolsky equation (\ref{teuk_eq2}), it may be assumed that both equations (\ref{rad_it}) and (\ref{ang_dt}) are derived from a single equation through the same process we have used to obtain equations (\ref{rad_e1}) and (\ref{ang_e1}) from Eq. (\ref{teuk_eq2}), i.e., using separation of variables and the transformation of coordinates (\ref{transf_1}). This reasoning leads to the equation:
\begin{multline}\label{tr_eqn}
\left(\frac{\partial}{\partial r}\left(X\frac{\partial}{\partial r}\right)+\frac{1}{\sin{\th}}\frac{\partial}{\partial\th}\left(\sin{\th}\frac{\partial}{\partial\th}\right)-\left(P(r)+a^2\cos^2{\th}\right)\frac{\partial^2}{\partial t^2}-\right.\\\left.2a\left(\frac{r-m}{\ve_{0}m}-\cos{\th}\right)\frac{\partial^2}{\partial t\partial \vp}-s^2\left(\frac{r-r_{+}}{r-r_{-}}+\frac{1-\cos{\th}}{1+\cos{\th}}\right)\right)e^{i(\om t+k\vp)}\bar{h}_{s}(r)\bar{v}_{s}(\th)=0,
\end{multline}
where $\ve_0=(r_{+}-r_{-})/(r_{+}-r_{-})$ and the function $P(r)$ is defined as follows:
\begin{equation}\label{P_func}
P(r)=4m^2(1+s_1^2+s_2^2)^2\frac{(r-r_{-})}{(r-r_{+})}+\frac{8m^2}{\vk}(1+s_1^2+s^2_2+2s^2_1s^2_2)(r-r_{-})+r^2-2mr.
\end{equation}
Equation (\ref{tr_eqn}) is in agreement with the corresponding equations in the Kerr case \cite{Whiting_JMP89}, and with the recently derived equation for the scalar field ($s=0$) in a more general background geometry with four distinct charges \cite{Cvetic_PRL20}. We also point out that the spin-dependent term in the equation (\ref{tr_eqn}) is of the same form as for the Kerr background, and supposedly similar dependence may take place for a more general background geometry. The only term in Eq. (\ref{tr_eqn}) which depends on the electric and magnetic charges is the function $P(r)$, similarly to what was observed in \cite{Cvetic_PRL20}. We also note that $P(r)$ is positive in the outer domain ($r>r_+$), which is very important for showing that there are no unstable modes.

Equation (\ref{tr_eqn}) allows us to derive the inverse auxiliary metric $\hat{g}^{\mu\nu}$ up to a conformal factor. Before we obtain this metric, we point out that Eq. (\ref{tr_eqn}) can be represented as a ``massive'' scalar field equation in terms of the auxiliary metric, namely: $\hat{\nabla}_{\mu}\hat{\nabla}^{\mu}\Phi-f(r,\th)\Phi=0$, where $\hat{\nabla}$ denotes the covariant differentiation operator with respect to the metric $\hat{g}$. Now we easily write components of the inverse auxiliary metric:
\begin{gather}
\nonumber\hat{g}^{tt}=-\frac{1}{\Omega^2}(P(r)+a^2\cos^2{\th}), \quad \hat{g}^{rr}=\frac{X}{\Omega^2},\\ \hat{g}^{\th\th}=\frac{1}{\Omega^2}, \quad \hat{g}^{t\vp}=-\frac{a}{\Omega^2}\left(\frac{r-m}{\ve_{0}m}-\cos{\th}\right).
\end{gather}
Imposing that the equation is exactly of the form (\ref{tr_eqn}) we obtain explicitly the conformal factor $\Omega$:
\begin{equation}
\Omega^2=a\sqrt{X}\left(\frac{r-m}{\ve_{0}m}-\cos{\th}\right)\sin{\th}.
\end{equation}
Finally, we write the auxiliary metric:
\begin{equation}\label{aux_metr}
d\hat{s}^2=\Omega^2\left(\frac{dr^2}{X}+d\th^2\right)+\frac{X}{\Omega^2}(P+a^2\cos^2{\th})\sin^2{\th}d\vp^2-2\sqrt{X}\sin{\th}dtd\vp.
\end{equation}
This metric is in agreement with the expression derived in \cite{Cvetic_PRL20} for the scalar case. The Killing vector of time translation $K^{\mu}=\partial/\partial t$ is everywhere null for the metric (\ref{aux_metr}). We will show that for the auxiliary metric (\ref{aux_metr}) the energy density current for the scalar field $J^{\mu}=K^{\la}T^{\mu}_{\la}$ is positive at least outside the black hole. As we have noted above, the equation (\ref{tr_eqn}) can be derived form a ``massive'' scalar field Lagrangian, which may be written in the form
\begin{equation}\label{scal_lagr}
{\cal L}=-\frac{1}{2}\hat{\nabla}_{\la}\Phi^{*}\hat{\nabla}^{\la}\Phi-\frac{f}{2}|\Phi|^2,
\end{equation}
where $f=s^2\left(\frac{r-r_{+}}{r-r_{-}}+\frac{1-\cos{\th}}{1+\cos{\th}}\right)$, that is positive in the outer region. The stress-energy tensor for the field $\Phi$ is
\begin{equation}\label{str_en}
T_{\mu\nu}=-\frac{1}{2}\left(\hat{\nabla}_{\mu}\Phi^{*}\hat{\nabla}_{\nu}\Phi+\hat{\nabla}_{\mu}\Phi\hat{\nabla}_{\nu}\Phi^{*}-\hat{g}_{\mu\nu}\left(\hat{\nabla}_{\la}\Phi^{*}\hat{\nabla}^{\la}\Phi+f|\Phi|^2\right)\right).
\end{equation}
Using the above relation for calculation of the energy current $J^{0}$ and taking into consideration the factor $\sqrt{-\hat{g}}$, we obtain:
\begin{equation}\label{current}
\sqrt{-\hat{g}}J^{0}=\frac{1}{2}\left((P(r)+a^2\cos^2{\th})\left|\partial_{t}\Phi\right|^2+X\left|\partial_{r}\Phi\right|^2+\left|\partial_{\th}\Phi\right|^2+s^2\left(\frac{r-r_{+}}{r-r_{-}}+\frac{1-\cos{\th}}{1+\cos{\th}}\right)|\Phi|^2\right).
\end{equation}
This current is manifestly positive outside the black hole. The current in (\ref{current})  generalizes the corresponding relation for the Kerr background \cite{Whiting_JMP89}. It is also a higher-spin generalization for the STU black hole \cite{Cvetic_PRL20} with pairwise-equal charges. Conservation of energy ${\cal E}=\int d^3{x}\sqrt{-\hat{g}} J^{0}$ together with positive definiteness of the integrand give rise to the conclusion that unstable modes should be withdrawn from the solution.

\section{Connection relations and mode stability for massless fermionic perturbations}

\indent Fermionic perturbations should be considered separately, because the differential part of the transformations is not applicable for a half-integer spin. Indeed, if $s\in\mathbb{Z}/2$ the parameter $n=s+m$, which was assumed to be integer as an order of differentiation, turns to be an half-integer as well. Here we examine spin $s=1/2$ fermion only and we use the Dirac equation instead of the general Teukolsky equation. In addition for spin $s=1/2$ fermions there is no superradiance, therefore the Whiting's transformation can be avoided. Recently it was shown that, to prove the absence of unstable modes in the Kerr case, some analytical properties of the differential equation and the so-called connection conditions, which relate the functions of opposite directions of spin, can be used \cite{Duztas_PRD16}. 

Throughout this derivation, we shall assume that the angular frequency $\omega$ is real. For the Kerr black hole, one may show that this case is possible through recourse to the work of Hartle and Wilkins~\cite{Hartle}. The basic reasoning goes as follows: i) The Schwarzschild black hole, which corresponds to the $a\to 0$ limit of Kerr, is known to be stable, so that the ingoing modes $Y^{(in)}$ cannot have zeros. ii) as  $a$ is continuously increased, a zero of $Y^{(in)}$ cannot move to the upper half of the complex plane without crossing the real axis. To rigorously prove the second assertion it is necessary  to show that $Y^{(in)}$, viewed as a function of $a$ and $\omega$, is analytic in these variables, so that there are no branch points and zeros cannot reach the upper half-plane without crossing the real axis. This was carefully proved in the aforementioned Ref.~\cite{Hartle}. For the purposes of this paper, we shall assume that this reasoning is still valid for the background at hand, which amounts to the assumption that the addition of the $\rm{U(1)}$ charges in the metric (which may be seem as a smooth deformation of the Kerr background) does not create branch points or otherwise affect the reasoning above.   

To derive the connection relations, we write two pairs of equations, for the radial and angular components of the Dirac spinor, which appear due to separation of variables \cite{Cvetic_arxiv2023}:
\begin{eqnarray}\label{rad_sys1}
\sqrt{X}\hat{\cal D}_{-}\bar{\cal R}_{\frac{1}{2}}(r)=\la \bar{\cal R}_{-\frac{1}{2}}(r),\\ \label{rad_sys2} \sqrt{X}\hat{\cal D}_{+}\bar{\cal R}_{-\frac{1}{2}}(r)=\la \bar{\cal R}_{\frac{1}{2}}(r),
\end{eqnarray}
where the differential operators $\hat{\cal D}_{\pm}=\partial_{r}\mp iK/F$, the functions $K(r)$ and $F(r)$ are defined above and $\la$ is the separation constant, related to the parameter $\la_{k,s}$ of the radial Teukolsky equation (\ref{mast_2}). Radial components of Dirac spinor are denoted as $\bar{\cal R}_{\pm\frac{1}{2}}(r)$ to avoid confusion with the radial wave functions of the Teukolsky equation (\ref{mast_2}). The radial wave function for $s=-\frac{1}{2}$ in the system (\ref{rad_sys1})-(\ref{rad_sys2}) is identical to the corresponding Teukolsky wave function (\ref{mast_2}), i.e. $\bar{\cal R}_{-\frac{1}{2}}(r)={\cal R}_{-\frac{1}{2}}(r)$.  To achieve complete agreement  for $s=\frac{1}{2}$ instead of the Dirac wave function $\bar{\cal R}_{\frac{1}{2}}$ we define a new function ${\cal R}_{\frac{1}{2}}(r)$ as follows: ${\cal R}_{\frac{1}{2}}(r)=X^{-\frac{1}{2}}\bar{\cal R}_{\frac{1}{2}}(r)$. Now the system (\ref{rad_sys1})-(\ref{rad_sys2}) can be rewritten as follows:
\begin{gather}\label{r_s11}
X\left(\hat{\cal D}_{-}+\frac{X'}{2X}\right){\cal R}_{\frac{1}{2}}(r)=\la {\cal R}_{-\frac{1}{2}}(r),\\ \label{r_s22} \hat{\cal D}_{+}{\cal R}_{-\frac{1}{2}}(r)=\la{\cal R}_{\frac{1}{2}}(r).
\end{gather}

The main advantage of the system (\ref{r_s11})-(\ref{r_s22}) is the fact that both wave functions ${\cal R}_{\pm\frac{1}{2}}(r)$ coincide with the corresponding radial Teukolsky wave functions (\ref{mast_2}). 

We also write the system for angular components of the spinor wave function:
\begin{gather}\label{ang_s1}
\hat{\cal L}_{+}{\cal S}_{\frac{1}{2}}(\th)=\la {\cal S}_{-\frac{1}{2}}(\th),\\ \label{ang_s2} \hat{\cal L}_{-}{\cal S}_{-\frac{1}{2}}(\th)=-\la {\cal S}_{\frac{1}{2}}(\th),
\end{gather}
where $\hat{\cal L}_{\pm}=\partial_{\th}\pm\left(a\om\sin{\th}+\frac{k}{\sin{\th}}\right)$ and ${\cal S}_{\pm\frac{1}{2}}(\th)$ denote corresponding angular components of the spinor wave function. If the frequency $\om$ is real then the differential operators $\hat{\cal L}_{\pm}$ is a real one. The angular components ${\cal S}_{\pm}(\th)$ are also a real function as a natural generalization of spin spherical harmonics. In this case the separation constant $\la$ is real as well. These considerations are crucial for the proof of absence of unstable modes.

To obtain the connection relations it is necessary to have asymptotic relations for the radial wave functions ${\cal R}_{\pm\frac{1}{2}}(r)$ near the horizon and at infinity. To achieve this aim, we use the Teukolsky equation (\ref{mast_2}) and, to make this analysis simpler, we introduce the tortoise coordinate $r_{*}$, defined as follows $dr_{*}=\frac{F(r)}{X}dr$. Now, at infinity $r\to\infty$ ($r_{*}\to+\infty$) the Teukolsky equation (\ref{mast_2}) takes the following approximate form:
\begin{equation}
\frac{\partial^2 {\cal R}_{s}}{\partial r_{*}^2}+\left(\om^2+\frac{2is\om}{r}\right){\cal R}_{s}\approx 0.
\end{equation}
This relation holds up to $\sim{\cal O}(1/r^2)$ terms, which definitely goes to zero if $r\to\infty$. The asymptotic solutions are derived easily and they can be written as follows:
\begin{equation}\label{R_asinf}
{\cal R}_{s}\simeq Y^{(in)}_{s}\frac{e^{-i\om r_{*}}}{r}+Y^{(out)}_{s}\frac{e^{i\om r_{*}}}{r^{2s+1}},
\end{equation}
where $Y^{(in)}_s$ and $Y^{(out)}_{s}$ are amplitudes for ingoing and outgoing waves respectively for the corresponding values of spin $s$.

Near the horizon ($r\to r_{+}$ or $r_{*}\to-\infty$) the Teukolsky equation can be rewritten in the form:
\begin{equation}
\frac{\partial^2 {\cal R}_{s}}{\partial r_{*}^2}+\left(\tilde{\om}-is\frac{r_{+}-m}{F(r_{+})}\right)^2{\cal R}_{s}\approx 0,
\end{equation}
 and here $\tilde{\om}=\om+\frac{ka}{F(r_{+})}$. The solutions near the horizon should be ingoing waves (infalling particles are examined), therefore simple analysis shows that they can be written in the form:
\begin{equation}\label{R_ashor}
{\cal R}_{s}(r)\simeq Z^{(in)}_{s}e^{-i\left(\tilde{\om}-is\frac{r_{+}-m}{F(r_{+})}\right)}\simeq Z^{(in)}_{s}X^{-\frac{s}{2}}e^{-i\tilde{\om}r_{*}}.
\end{equation}
Here we point out that our asymptotic relations for the radial wave functions ${\cal R}_{s}(r)$ at the horizon and at the infinity are in perfect agreement with the corresponding relations obtained for the Teukolsky equation in Kerr case \cite{Teukolsky_ApJ1973}.

Now, using the asymptotic relations (\ref{R_asinf}) and (\ref{R_ashor}) and the equations (\ref{r_s11}) and (\ref{r_s22}), we find the connection relations between the amplitudes $Y^{(in)}_{s}$, $Y^{(out)}_{s}$ and $Z^{(in)}_{s}$ for opposite values of spin. Taking the asymptotic relation (\ref{R_asinf}) and the equations (\ref{r_s11}) and (\ref{r_s22}) we derive respectively:
\begin{equation}\label{conn_1}
2i\om Y^{(out)}_{\frac{1}{2}}=\la Y^{(out)}_{-\frac{1}{2}}, \quad -2i\om Y^{(in)}_{-\frac{1}{2}}=\la Y^{(in)}_{\frac{1}{2}}.
\end{equation}
Finally, if we take the asymptotic relations (\ref{R_ashor}) and the equation (\ref{r_s22}) we arrive at the following relation:
\begin{equation}\label{conn_2}
Z^{(in)}_{-\frac{1}{2}}\left(r_{+}-m-2i\tilde{\om}F(r_{+})\right)=\la Z^{(in)}_{\frac{1}{2}}.
\end{equation}
A similar connection relation for the Kerr background was obtained in \cite{Duztas_PRD16} and our relations are consistent with it. 

The other type of connection relations associates the amplitudes $Y^{(in)}_{s}$, $Y^{(out}_{s}$ and $Z^{(in)}_{s}$ at infinity and at the horizon for the same orientation of spin. To obtain this connection relation we rewrite the Teukolsky equation in the form:
\begin{equation}\label{diff_eW}
\frac{\partial^2 {\cal X}_{s}}{\partial r_{*}^2}+U(r,s){\cal X}_{s}=0,
\end{equation}
where derivatives are taken with respect to the tortoise coordinate $r_{*}$, ${\cal X}_s=X^{\frac{s}{2}}F^{\frac{1}{2}}{\cal R}_s(r)$ and the explicit form of the function $U(r,s)$ can be calculated, but it is not important for our following considerations. It is known that for any differential equation of the form (\ref{diff_eW}), the Wronskian is a constant, namely for a pair of its two independent solutions ${\cal X}^{(1)}_{s}$ and ${\cal X}^{(2)}_{s}$ we can write:
\begin{equation}
W({\cal X}^{(1)}_{s},{\cal X}^{(2)}_{s})\equiv{\cal X}^{(1)}_{s,r_{*}}{\cal X}^{(2)}_{s}-{\cal X}^{(1)}_{s}{\cal X}^{(2)}_{s,r_{*}}=const,
\end{equation}
where the derivatives are taken with respect to $r_{*}$. The potential $U(r,s)$ satisfies $U^{*}(r,s)=U(r,-s)$ while all the other parameters of the potential $U(r,s)$ are held fixed, therefore the invariance of the Wronskian can be rewritten in the form:
\begin{equation}\label{Wr_cond}
W({\cal X}_{\frac{1}{2}},{\cal X}^{*}_{-\frac{1}{2}})_{r_{*}}=W({\cal X}_{\frac{1}{2}},{\cal X}^{*}_{-\frac{1}{2}})_{\infty}.
\end{equation}
Using the explicit relation for the function ${\cal X}_{s}$ and the asymptotic relations (\ref{R_asinf}) and (\ref{R_ashor}) we can calculate both sides of latter relation. Namely at the horizon we obtain:
\begin{equation}\label{Wr_h}
W({\cal X}_{\frac{1}{2}},{\cal X}^{*}_{-\frac{1}{2}})_{r_{*}}=Z^{(in)}_{\frac{1}{2}}{Z^{(in)}_{-\frac{1}{2}}}^{*}\left(-(r_{+}-m)-2i\tilde{\om}F(r_{+})\right),
\end{equation}
and at infinity we arrive at the following form:
\begin{equation}\label{Wr_inf}
W({\cal X}_{\frac{1}{2}},{\cal X}^{*}_{-\frac{1}{2}})_{\infty}=2i\om\left(Y^{(out)}_{\frac{1}{2}}{Y^{(out)}_{-\frac{1}{2}}}^{*}-Y^{(in)}_{\frac{1}{2}}{Y^{(in)}_{-\frac{1}{2}}}^{*}\right).
\end{equation}
Finally, substituting the connection relations (\ref{conn_2}) and (\ref{conn_1}) into the relations (\ref{Wr_h}) and (\ref{Wr_inf}) respectively and equating both of them we obtain:
\begin{equation}
|Z^{(in)}_{\frac{1}{2}}|^2=|Y^{(in)}_{\frac{1}{2}}|^2-\frac{4\om^2}{\la^2}|Y^{(out)}_{\frac{1}{2}}|^2.
\end{equation}
 We also point out that to write the upper equation we have used our assumption  that the frequency $\om$ and the separation constant $\la$ are real. Clearly, if $Y^{(in)}_{\frac{1}{2}}=0$, the above equation does not have any solution corresponding to real $\omega$. This means that a zero of  $Y^{(in)}_{\frac{1}{2}}$ could never move to the upper half-plane by crossing the real axis. As reasoned above, this means that, if $Y^{(in)}_{\frac{1}{2}}$ is an analytic function of $a$ and $\omega$, it cannot develop a zero and cannot therefore give rise to an unstable mode. A similar connection relation can be written for $s=-\frac{1}{2}$ and the exact same  argument may be conducted to rule out unstable perturbations.
\section{Conclusions}
\indent The mode stability for black holes has become a very active area of study in recent years among both mathematicians' and physicists' communities \cite{Kerr_stability,Cvetic_PRL20,Cvetic_PRD22,Casals_CMP22}. To examine mode stability for General Relativistic backgrounds, the celebrated Teukolsky equation is often used \cite{Teukolsky_ApJ1973,Whiting_JMP89}, but for more general black hole backgrounds, such as those from  String Theory or/and Supergravity, an equation of this kind is not yet known. 

In a previous work from two of the authors \cite{Cvetic_arxiv2023}, the separability of the Dirac equation on a rotating pair-wise equal charged STU black hole background has been analyzed. The results from this investigation, as well as earlier ones concerning scalar fields \cite{Cvetic_PRL20}, have led us to conjecture the Teukolsky-like equation that has been used in this work to study minimally coupled probe fields of higher spins. Thus, we have used equation (\ref{teukolsky_eq1}) to derive decoupled equations for the radial and angular parts, obtained though the method of separation of variables. Generalizing Whiting's procedure, we study mode stability for minimally coupled fields of higher integer spin. The obtained results are in agreement with the analysis conducted in Ref.~\cite{Cvetic_PRL20} for scalar fields in a more general metric.  There are some subtleties for the fermionic case, which preclude a straightforward application of Whiting's procedure. To examine mode stability for the torsion-modified massless Dirac equation, we derive connection relations relating the two components of this equation, following the approach considered in \cite{Duztas_PRD16}. The connection relations are shown to be in direct contradiction with the existence of unstable modes in the theory, which may thus be ruled out. We note that spin $1/2$ fields were observed to cause instability of the event horizon \cite{Duztas_PRD16}, but this issue needs careful study and will be examined elsewhere. 

An important direction for further study is the application of the developed techniques for five dimensional STU black holes, where up to now only scalar field mode stability has been examined \cite{Cvetic_PRD22}. 

\section*{Acknowledgments}
{We would like to thank Gary Gibbons, Chris Pope, Bernard Whiting and Haoyu Zhang for discussions on topics related to this paper. MC is partially supported by the Slovenian Research Agency (ARRS No. P1-0306) and Fay R. and Eugene L. Langberg Endowed Chair funds, by DOE Award (HEP) DE-SC0013528, by a University Research Foundation Grant at the University of Pennsylvania and by the Simons Foundation Collaboration grant $\# 724069$. MMS was partially supported by the Fulbright Program grant for visiting scholars. MAL was funded by Conselho Nacional de Desenvolvimento Cientifico e Tecnologico (CNPq), grant No. 401991/2022-9 and Paraiba State Research Foundation (FAPESQ-PB) grant No. 0015/2019.} 

\end{document}